\documentclass{nic-series}

\begin{document}                                                                

\title{ Numerical simulation of high quark densities
        in QCD with two colours }

\author{ I.\ Montvay \inst{1}   \and
         S.\ Hands \inst{2}     \and
         L.\ Scorzato \inst{2}  \and
         J.I.\ Skullerud \inst{1} }

\institute{ Deutsches Elektronen-Synchrotron DESY \\
            Notkestrasse 85, D-22603 Hamburg, Germany
            \and
            Department of Physics, University of Wales Swansea \\
            Singleton Park, Swansea SA2 8PP, U.K. }

\newcommand{\beq}{\begin{equation}}
\newcommand{\eeq}{\end{equation}}
\newcommand{\half}{\frac{1}{2}}                                                 
\newcommand{\rar}{\rightarrow}                                                  
\newcommand{\lar}{\leftarrow}
\newcommand{\LCB}{\raisebox{-0.3ex}{\mbox{\LARGE$\left\{\right.$}}}
\newcommand{\RCB}{\raisebox{-0.3ex}{\mbox{\LARGE$\left.\right\}$}}}
\newcommand{\U}{\mathrm{U}}
\newcommand{\SU}{\mathrm{SU}}
                                                                                
\maketitle

\begin{abstracts}
 The DESY-Swansea Collaboration performed numerical simulations
 investigating SU(2) lattice gauge theory at non-zero chemical potential
 with one staggered quark flavour in the adjoint representation.
 This lattice model has similar features to QCD itself and its study
 gives interesting insights into some open problems of high density quark
 matter.
 In particular the r\^ole of the ``sign problem'' can be clarified in
 connection with diquark condensation and the phase diagram.
\end{abstracts}       

\section{Introduction}\label{sec1}

 The fundamental theory of strongly interacting matter is Quantum
 Chromo-Dynamics (QCD), which is a relativistic quantum field theory
 with quarks and gluons as quanta of the elementary constituent
 fields.
 Physical properties of strongly interacting matter, manifested in
 high energy heavy ion collisions, in the early Universe or in neutron
 stars, can, in principle, all be deduced from this theory.
 An important calculational method in QCD is numerical Monte Carlo
 simulation on space-time lattices.

 In recent years, significant progress has been made in understanding
 the phase diagram of QCD at high quark- (or baryon-) number density.
 On the basis of model calculations \cite{INSTANTON,MEANFIELD}, it is
 now believed that the ground state of QCD at high density and low
 temperatures is characterised by a diquark condensate which
 spontaneously breaks gauge and/or baryon number symmetries (for recent
 reviews see \cite{REVIEW}).
 This leads to phenomena such as {\em colour superconductivity} and/or
 {\em colour-flavour locking} which have substantial influence on the
 physics of high density hadronic (quark-) matter and have a major
 impact, for instance, on the properties of heavy neutron star cores.
 However, although the results appear to be qualitatively independent
 of the specific model and approximation employed, little can be said
 quantitatively due to the lack of a first-principles, non-perturbative
 method that can access the relevant regions of the phase diagram.

 Non-perturbative Monte Carlo simulations in lattice QCD, which
 are first-principles methods, fail when a chemical potential term
 $\mu n_B$, where $n_B$ is the quark number density, is added to the
 Lagrangian.
 The reason is that the Euclidean-space fermion determinant becomes
 complex.
 As a consequence, standard simulation algorithms cannot be applied.
 However, it is possible to study QCD-like theories where the fermion
 determinant remains real even at non-zero $\mu$.  
 These theories can be used as testbeds to examine the validity of the
 models used to study real QCD, as well as directly to improve our
 understanding of phenomena such as diquark condensation and phase
 transitions in dense matter.
 Examples of such theories with real quark determinant are two-colour
 (SU(2) colour) QCD and QCD with adjoint quarks.

 The DESY-Swansea Collaboration studied two-colour QCD with staggered
 quarks in the adjoint representation \cite{ADJOINT,BANGALORE,DIQUARK}.
 The use of a particular lattice fermion formulation (with staggered
 fermions) is important because for real or pseudoreal representations
 of the gauge group the pattern of symmetry breaking is expected to
 be different in the continuum and at non-zero lattice spacing
 \cite{HKLM}. 
 In this QCD-like model it is possible to investigate several
 interesting questions.
 Its basic features are:
\begin{itemize}
\item
 For an odd number of staggered flavours $N$, the fermion determinant
 may be negative. 
 This means that this model has a {\em sign problem}, which
 may make simulations at large $\mu$ difficult.
 At the same time a sign is simpler than the continuum of phases in QCD.
 It may thus be feasible to make progress using standard means, or at
 least expose physical distinctions between the positive and negative
 sectors.
\item
 For zero quark mass ($m$) and vanishing chemical potential
 $m=\mu=0$ the $\U(N) \times \U(N)$ flavour symmetry is enhanced
 to a $\U(2N)$ symmetry which relates quarks to antiquarks. 
 This symmetry is broken by the chiral condensate, leaving $N(2N-1)$
 massless Goldstone modes, which become degenerate pseudo-Goldstone
 states for $m\neq0$.
\item
 When $N\geq2$ these pseudo-Goldstone states include gauge invariant
 scalar diquarks which, though degenerate with the pion at
 $\mu=0$, carry baryon number.
 These models with $N\geq2$ can be studied for $\mu\not=0$ by 
 chiral perturbation theory ($\chi$PT) \cite{KSTVZ}.
 The main result is that for $\mu>m_\pi/2$, where $m_\pi$ is the pion
 mass, the chiral condensate rotates into a diquark condensate (the two
 being related by a $\U(2N)$ rotation) while the baryon density
 increases from zero. 
\item
 The model with $N=1$ is not expected to contain any diquark
 pseudo-Goldstones and is not accessible to $\chi$PT. 
 One expects an onset transition as some $\mu_o\approx m_b/n_q>m_\pi/2$,
 where $m_b$ is the mass of the lightest baryon and $n_q$ its baryon
 charge. 
\item
 For $N=1$ the gauge invariant scalar diquark operator is forbidden by
 the Pauli Exclusion Principle; there is, however, a possibility of a
 gauge non-singlet, and hence colour superconducting, diquark condensate
 at large chemical potential.
\end{itemize}

 The numerical simulation of lattice gauge theories is performed on a
 hypercubic lattice in four dimensional Euclidean space.
 The fourth coordinate, besides the three space coordinates, is
 imaginary time.
 In the path integral formulation the Euclidean action is needed which
 gives the weight of lattice field configurations.
 For the gauge field part the standard Wilson action, weighted by the
 inverse gauge coupling $\beta$, is taken \cite{MM}.
 The fermionic part of the lattice action, with quark mass $m$ and
 chemical potential of quark charge $\mu$, can be defined as follows:
\begin{eqnarray}
S &=& \sum_{x,y}\bar\chi^p(x)D_{x,y}[U,\mu]\chi^p(y)+
m\bar\chi^p(x)\delta_{x,y}\chi^p(y)
\nonumber \\
&\equiv& \sum_{x,y}\bar\chi^p(x)M_{x,y}[U,\mu]\chi^p(y),
\label{eq01}
\end{eqnarray}
 where the index $p$ runs over $N$ flavours of staggered quark, 
 and $D$ is given by
\begin{eqnarray}
D_{x,y} &=& \half\sum_{\nu\not=0}\eta_\nu(x)
\left(U_\nu(x)\delta_{x,y-\hat\nu}
        -U_\nu^\dagger(y)\delta_{x,y+\hat\nu}\right)
\nonumber \\
&+& \half\eta_0(x)\left(e^\mu U_0(x)\delta_{x,y-\hat0}
        -e^{-\mu}U_0^\dagger(y)\delta_{x,y+\hat0}\right).
\label{eq02}
\end{eqnarray}
 The $\chi, \bar\chi$ are single spin component Grassmann objects, 
 and the phases $\eta_\mu(x)$ are defined to be
 $(-1)^{x_0+\cdots+x_{\mu-1}}$.
 The link matrices in the adjoint representation $U_\mu$ are real
 $3\times3$ orthogonal matrices.
 The Grassmann variables in the path integral can be integrated out,
 resulting in the determinant of the fermion matrix $M$ appearing in
 (\ref{eq01}).
 This {\em fermion determinant} takes into account the effects of
 virtual fermion-antifermion pairs on the gauge field.

 The main difficulty in numerical simulations of gauge theories with
 fermions is to include the fermion determinant.
 We have used two different simulation algorithms, the hybrid Monte
 Carlo (HMC) algorithm \cite{DKPR} which is the standard choice for QCD,
 and a Two-Step Multi-Bosonic (TSMB) algorithm \cite{TSMB} which has
 been developed recently in connection with a study of supersymmetric
 Yang-Mills theory.
\begin{figure}[htb]
\begin{center}
\epsfig{file=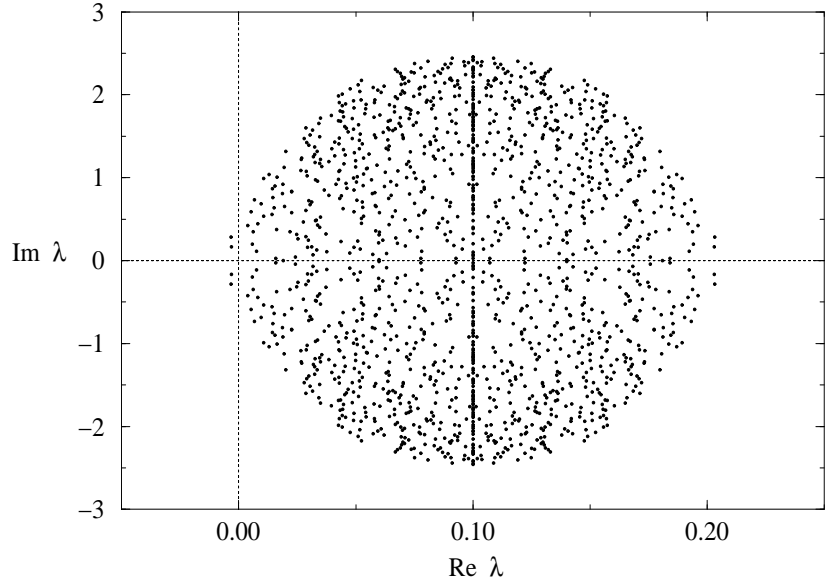,
        width=11cm,height=8cm,angle=0}
\caption{\label{fig01}\em
 Typical eigenvalue spectrum of the fermion matrix on a $4^3 \cdot 8$
 lattice at ($\beta=2.0,\; m=0.1,\; \mu=0.35$).}
\end{center}
\end{figure}

\section{The sign of the determinant}\label{sec2}

 At zero chemical potential $\mu=0$ the complex eigenvalues of the fermion
 matrix $M$ in (\ref{eq01}) lie on a line parallel to the imaginary axis
 with real part equal to the quark mass ${\mathrm Re}\lambda = m$.
 Since the eigenvalues come in complex conjugate pairs, the determinant
 of the fermion matrix at $\mu=0,\; m>0$ is positive.
 After introducing a non-zero chemical potential the eigenvalues
 spread out into an elliptical region of the complex plane (see, for
 instance, figure \ref{fig01}).

\begin{figure}
\begin{center}
\includegraphics[angle=-90,width=11cm]
{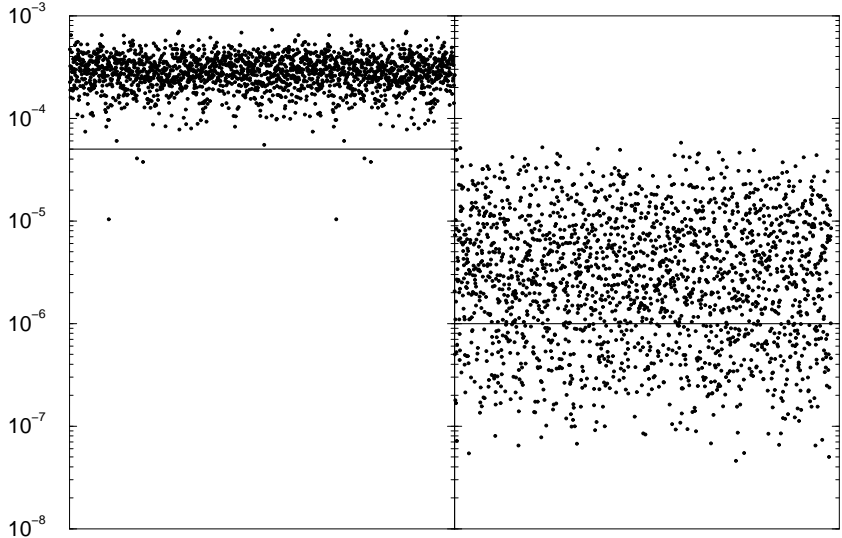}
\caption{\em
 The lowest eigenvalues of the hermitean fermion matrix, for
 $\mu=0.3$ (left) and $\mu=0.4$ (right).  The lines indicate the lower
 bound $\epsilon$ of the polynomial approximation employed.}
\label{fig02}
\end{center}
\end{figure}

 At large enough chemical potential the region occupied by the
 eigenvalues of the fermion matrix reaches the origin of the complex
 plane (this is actually the situation in figure \ref{fig01}), which
 has important consequences for the behaviour of the system.
 From the technical point of view this is manifested by the appearence
 of very small eigenvalues of the squared fermion matrix
 (see figure \ref{fig02}), which makes the numerical simulation
 difficult.
 The very small eigenvalues occur because there are sign changes of
 the fermion determinant.
 It turns out that this can only be seen in the TSMB simulation
 because the HMC algorithm is based on small steps in configuration
 space and cannot in practice change the determinant sign.
 This is an important observation showing that under these
 circumstances the standard HMC algorithm is not applicable because
 it is not ergodic.
 As we shall see below, the HMC simulation is very similar to
 the restriction of the TSMB simulation to the positive determinant
 sector.
 The omission of the gauge configurations with negative determinants
 has important consequences for the physical results.

\begin{figure*}
\begin{center}
\includegraphics[width=0.3\textwidth]
{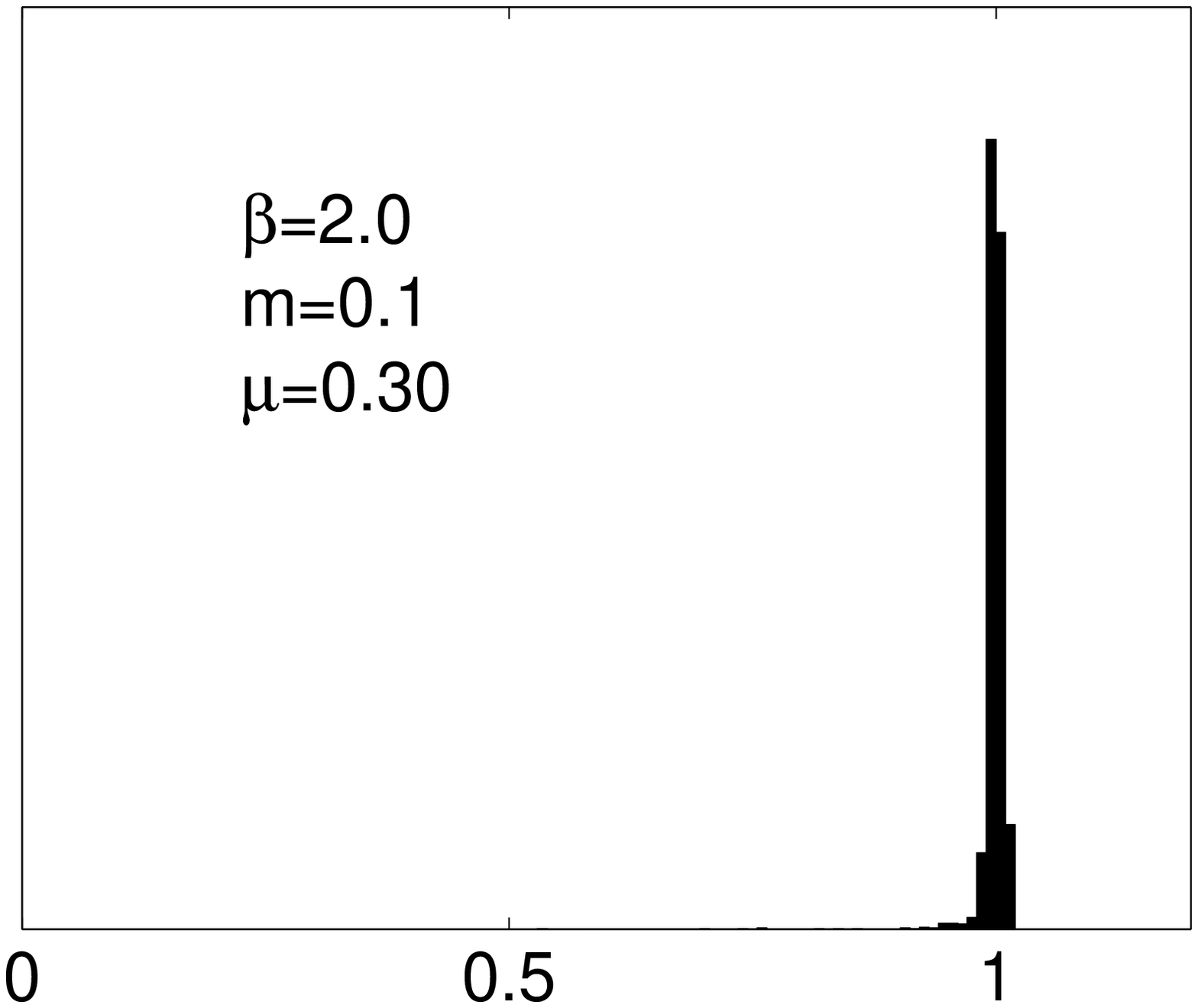}
\includegraphics[width=0.329\textwidth]
{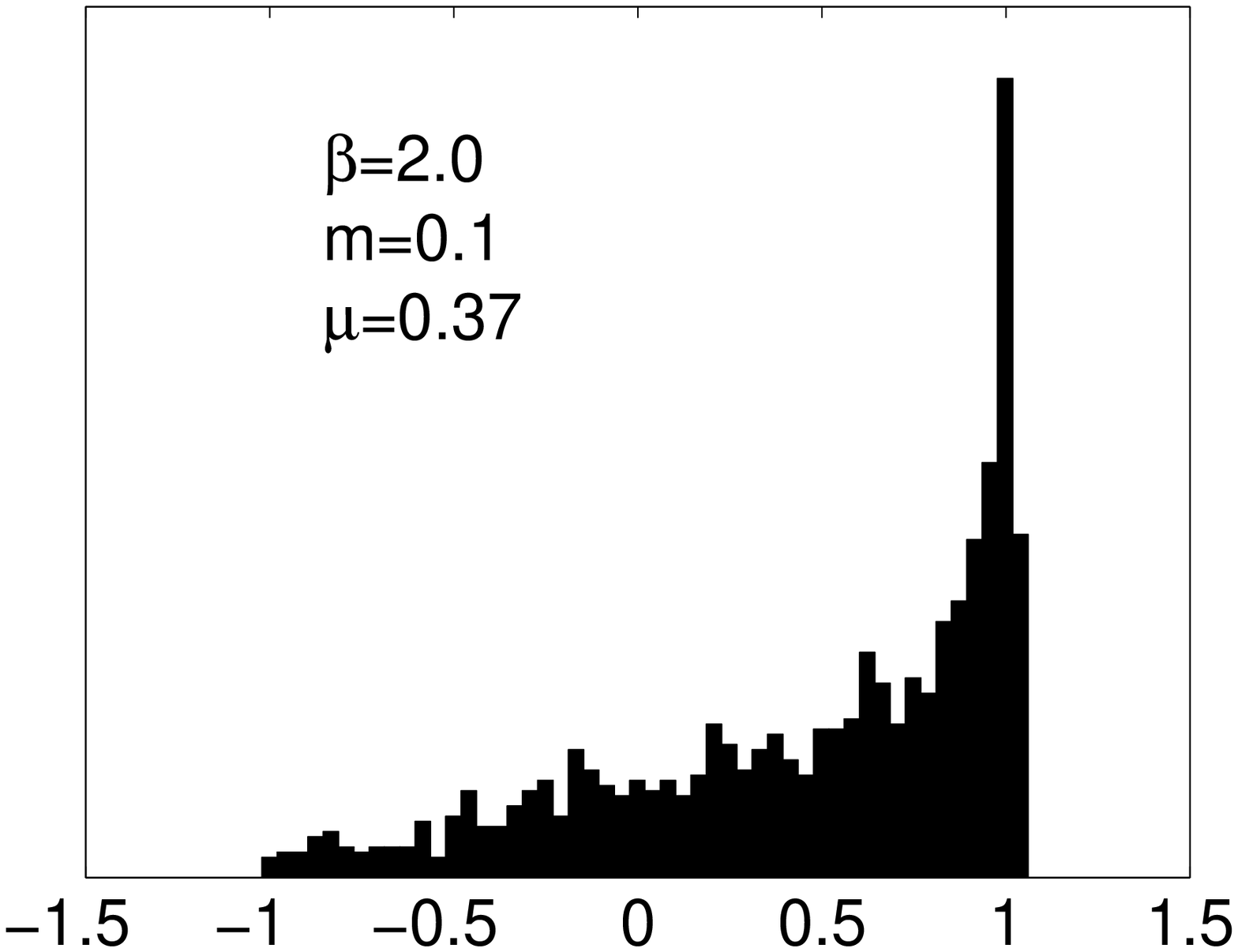}
\includegraphics[width=0.329\textwidth]
{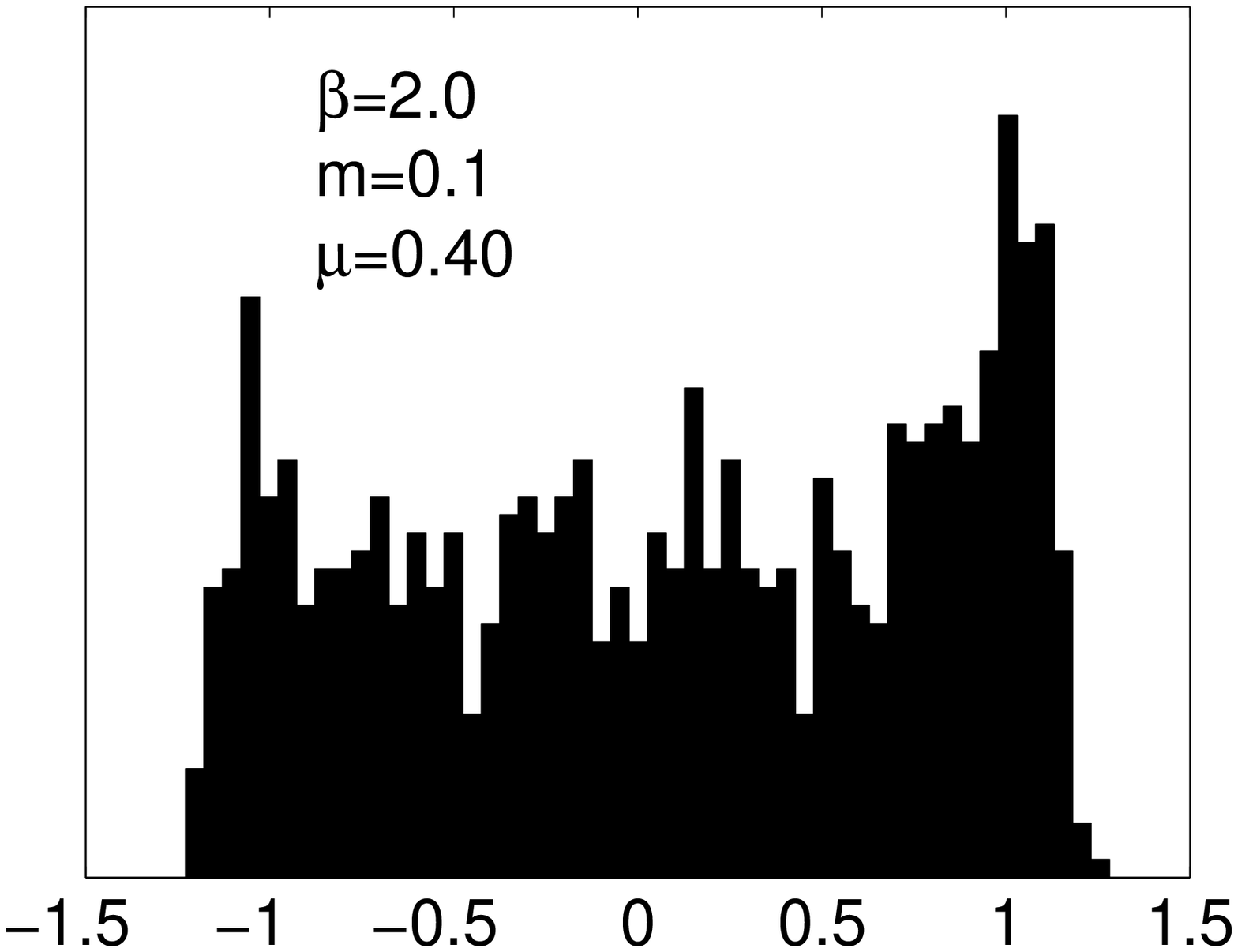}
\caption{\em
 Reweighting factors, for $\mu=0.3$ (left), $\mu=0.37$
 (middle) and $\mu=0.4$ (right).  As the proportion of negative
 determinant configurations increases, the preferred distribution of
 reweighting factors changes from being sharply peaked around 1 to
 nearly flat.  The polynomial orders $(n_1,n_2)$ are (48,500), (80,800)
 and (100,1000) respectively.}
\label{fig03}
\end{center}
\end{figure*}

 In simulations with the TSMB algorithm the qualitative change
 for increasing chemical potential can also be clearly seen in
 the distribution of the {\em reweighting factors} (see figure
 \ref{fig03}).
 These are required because the polynomial approximations applied
 in this algorithm \cite{POLYNOM} are not exact near zero eigenvalues.
 The reweighting factors, which also include the sign of the fermion
 determinant, are correcting for this.
 As it can be seen in figure \ref{fig03}, the reweighting factors are
 strongly peaked near 1 for a $\mu$ value where the spectrum does not
 yet touch zero ($\mu=0.3$).
 At $\mu=0.37$, where the spectrum just reaches the origin, some
 configurations with negative reweighting factor (i.e. negative
 determinant) start to appear.
 Finally, at $\mu=0.4$ the frequency of both signs is already almost
 equal and the sign problem is serious.

\section{Simulation results}\label{sec3}

\begin{figure}[htb]
\begin{center}
\includegraphics[width=11cm]
{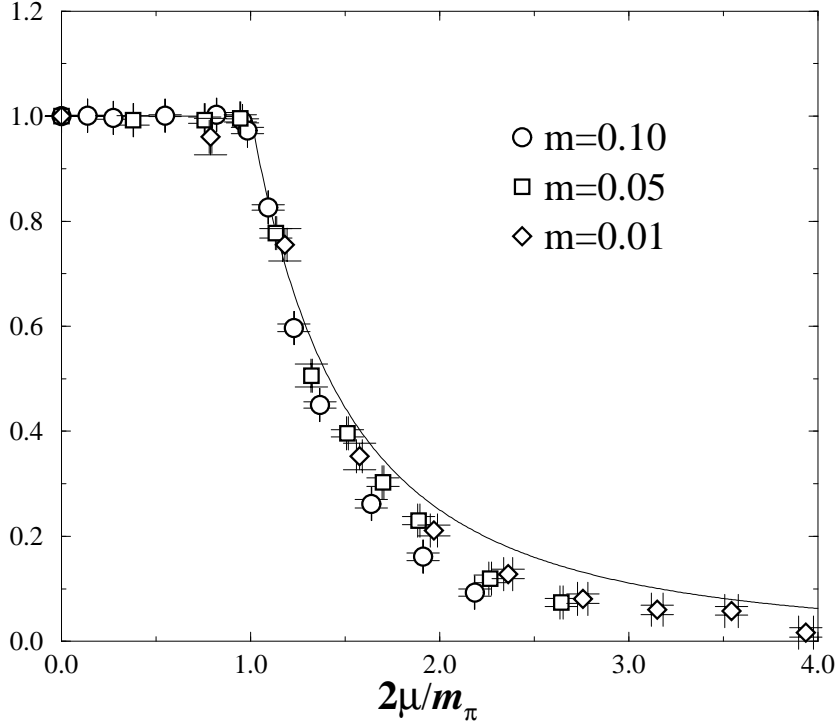}
\caption{\em
 Rescaled chiral condensate vs.\ chemical potential obtained in HMC
 simulations together with the $\chi$PT predictions.
\label{fig04}}
\end{center}
\end{figure}

 Having a suitable simulation algorithm (TSMB) which samples both
 positive and negative determinant sectors properly one can compare
 results with and without taking into account the determinant sign.
 Most of the results of the DESY-Swansea Collaboration have been
 obtained at $\beta=2.0,\; m=0.1$ on a $4^3 \cdot 8$ lattice
 \cite{ADJOINT,BANGALORE,DIQUARK}.
 It turned out that the results in the positive sector of the TSMB
 simulations are close to the HMC simulation results and both are
 qualitatively similar to the results of chiral perturbation theory for
 theories with diquark Goldstone modes (for instance, $N=2$ flavours)
 \cite{KSTVZ}.
 This is not surprising because taking the absolute value of the
 fermion determinant in the path integral measure is equivalent
 to consider $\sqrt{\det(M^\dagger M)}$ instead of $\det(M)$.
 An example of the good agreement between the results of HMC
 simulations and chiral perturbation theory predictions is shown
 by figure \ref{fig04} in case of the chiral condensate
 $\langle\bar{\psi}\psi\rangle$.

\begin{figure}
\begin{center}
\includegraphics[height=11cm,angle=-90]
{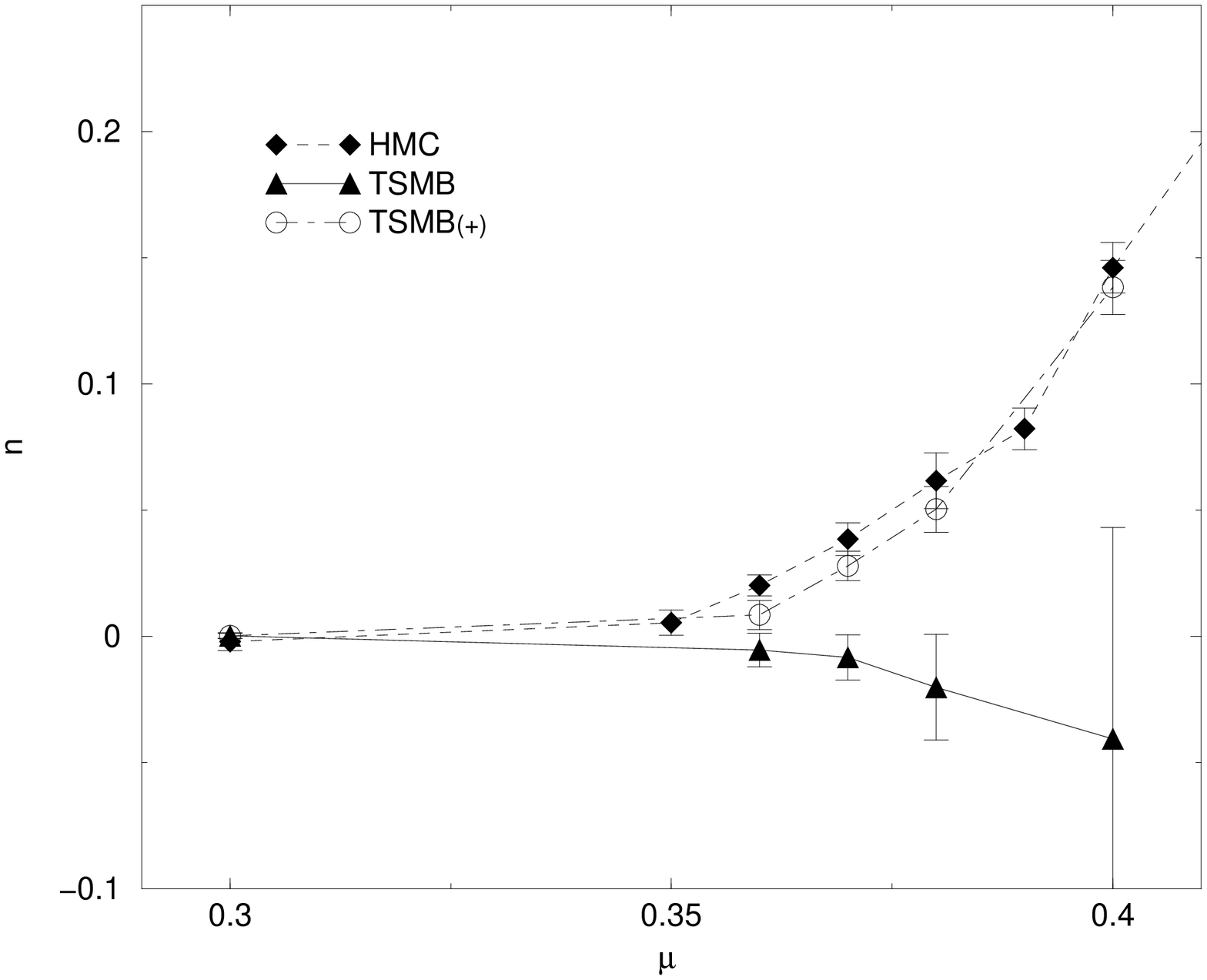}
\caption{\em
 The fermion density from TSMB and HMC simulations.
 In case of TSM simulations the result in the positive determinant
 sector is shown separately (open symbols).
}
\label{fig05}
\end{center}
\end{figure}

 The inclusion of the fermion determinant sign, in the single staggered
 flavour ($N=1$) model we are interested in, changes the physical
 results qualitatively.
 The main effect is that the phase transition at chemical potential
 $\mu \simeq \half m_\pi$, which is signalled for instance by the
 sharp drop of the chiral condensate in figure \ref{fig04}, disappears
 due to a cancellation between the contributions of positive and
 negative sectors.
 As an example one can consider the behaviour of the quark density
 shown in figure \ref{fig05}.
 The increase beyond $\mu=0.35$ disappears once the signs are taken
 into account properly.
 Unfortunately, the severe sign problem does not allow us to go much
 beyond $\mu=0.4$ where the onset phase transition of the $N=1$ model
 is expected.

 A natural explanation of the agreement between the HMC and positive
 determinant results and $\chi$PT predictions is
 that the positive determinant sector mimics a theory with two flavours:
 one quark and one `conjugate quark' \cite{STEPHAN}. 
 We may cast further light on this by studying the behaviour of the
 diquark condensate of the $N=2$ diquark operator 
\begin{equation}\label{eq03}
qq_{\bf3}=\frac{i}{2}\left[\chi^{p\,tr}(x)\varepsilon^{pq}\chi^q(x)+
\bar\chi^p(x)\varepsilon^{pq}\bar\chi^{q\,tr}(x)\right],
\end{equation}
 where $p,q=1,2$ are explicit flavour indices and $\varepsilon$ the
 antisymmetric tensor.
 A non-zero expectation value $\langle qq_{\bf3} \rangle \ne 0$ is
 expected in the two-flavour theory, violating the symmetry of the
 Lagrangian under global U(1) baryon number rotations.
 Physically this implies a superfluid ground state and (via
 Goldstone's theorem) exactly massless scalar excitations.
 This accounts for the proliferation of small eigenvalues observed in
 section \ref{sec2}. 
 The superfluid condensate may be determined by introducing a diquark
 source term in the action, which now describes two flavours,
\begin{equation}\label{eq04}
S[j] = S + j\sum_x qq_{\bf3}(x) \, .
\end{equation}
 One has to extract the condensate 
 $\langle qq_{\bf3}(j)\rangle\!=\!V^{-1}\partial\ln Z[j]/\partial j$,
 and extrapolate the results to $j=0$ \cite{MH}.
 The HMC simulations do support $\langle qq_{\bf3} \rangle \ne 0$
 for $\mu \geq m_\pi$, as expected.
 In the TSMB simulations taking into account the determinant sign
 the diquark condensate $\langle qq_{\bf3} \rangle$ is suppressed.

\section{Conclusions}\label{sec4}

 We have studied two-colour QCD with one flavour of staggered quark in
 the adjoint representation.
 We have employed two different simulation algorithms, and have gained
 insight into the optimal tuning of the TSMB algorithm in the high
 density regime.
 This is the preferred algorithm not only because it is capable of
 maintaining ergodicity via its ability to change the determinant sign, 
 but also because it more effectively samples small eigenmodes, which
 are important in the presence of a physical Goldstone excitation. 
 We find that the positive determinant sector behaves like a two-flavour
 model, and exhibits good agreement with chiral perturbation theory
 predictions for such a model.  
 At higher chemical potentials there are some indications of a breakdown
 of $\chi$PT, and possible signs of a further phase transition.
 However, data from larger volumes and smaller bare quark masses 
 would be needed to make these observations definitive. 

 Above the onset transition in the positive determinant sector, we have
 successfully obtained a signal for a non-zero two-flavour diquark
 condensate $\langle qq_{\bf3}\rangle$, indicating a superfluid ground
 state for $\mu>m_\pi/2$.
 The chiral condensate rotates into this diquark condensate, in good
 quantitative agreement with the behaviour predicted by $\chi$PT. 
 This feature also enabled us to achieve reasonable control over the
 necessary $j\to0$ extrapolation \cite{DIQUARK}.

 When the negative determinant configurations are included in the
 measurement, the onset transition and diquark condensation disappear.
 This is what we would expect for the one-flavour model and is consistent
 with the Pauli Exclusion Principle. 
 There is a strong evidence for this scenario, providing a conclusive
 demonstration, should any still be needed, that the determinant 
 sign plays a decisive role in determining the ground state of systems
 with $\mu\not=0$.
 Unfortunately, the severity of the sign problem means we have not been
 able to locate the real onset transition for this model.



\end{document}